\documentclass[useAMS,usenatbib]{mn2e}

 \usepackage{journal_abbr_mac,graphicx}

\title[On calibration of some distance scales in astrophysics]{On calibration of some distance scales in astrophysics}
\author[B. Vukoti\'c, M. Jurkovi\'c, D. Uro\v sevi\'c and B. Arbutina]{B. Vukoti\'c$^{1}$\thanks{E-mail:
bvukotic@aob.rs (BV -- corresponding author), mojur@aob.rs (MJ), dejanu@math.rs (DU), arbo@math.rs (BA).}, M. Jurkovi\'c$^{1}$, D. Uro\v sevi\'c$^{2,3}$, and B. Arbutina$^{2}$\\
$^{1}$Astronomical Observatory, Volgina 7, 11060 Belgrade 38, Serbia\\
$^{2}$Department of Astronomy, Faculty of Mathematics, University of Belgrade, Studentski trg 16, 11000 Belgrade, Serbia\\
$^{3}$Isaac Newton Institute of Chile, Yugoslavia Branch\\
}
\begin{document}

\date{Accepted -- -- --. Received -- -- --; in original form \today}


\maketitle

\label{firstpage}

\begin{abstract}
We present a method for distance calibration without using
standard fitting procedures. Instead we use random resampling to
reconstruct the probability density function (PDF) of calibration
data points in the fitting plane. The resulting PDF is then used
to estimate distance-related properties. The method is applied to
samples of radio surface brightness to diameter ($\Sigma-D$) data
for the  Galactic supernova remnants (SNRs) and planetary nebulae
(PNe), and period-luminosity ($PL$) data for the Large Magellanic
Cloud (LMC) fundamental mode classical Cepheids. We argue that
resulting density maps can provide more accurate and more reliable
calibrations than those obtained by standard linear fitting
procedures. For the selected sample of the Galactic SNRs, the
presented PDF method of distance calibration results in a smaller
average distance fractional error of up to $\approx16$ percentage
points. Similarly,  the fractional error is smaller for up to
$\approx8$ and $\approx0.5$ percentage points, for the samples of
Galactic PNe and LMC Cepheids, respectively. In addition, we
provide a PDF-based calibration data for each of the samples.
\end{abstract}


\section{Introduction}

Scaling relations are widely used, sometimes as the only option,
to determine the relevant properties of astrophysical objects. One
particularly important property in astrophysical studies is the
distance to a particular object. Direct measurement of distances
is often not possible, and the only way to infer the distance is
from a scaling relation. The calibration of these relations is
very important and extensive scientific efforts have been made to
assess the quality of the calibration data samples and applied
calibration procedures. Most commonly, sample of calibrators is
fitted with some analytical functional dependence, where one of
the data variables is dependent on one or more remaining data
variables. In cases where all data variables have significant
uncertainties and cannot be resolved to dependent/independent
ones, fitting procedures are adjusted accordingly. If there are
$N$ calibrators in a particular calibrating sample, with $n$
coordinates per data point, then there are $ n \times N $ numbers
of information in that particular calibrating sample. When
fitting, all of this information is projected into the linear fit
parameters. The initial information contained in the calibration
sample is thus reduced and averaged out. While this might not be
problematic for the samples with strong functional dependence, in
the case of loose correlations (often used in astrophysics) it
might cause a significant difference, depending on the type of
functional dependence, type of offsets from the best fit line and
other assumptions of the applied fitting procedure
\citep{Isobe_etal_1990ApJ, urosevic_etal_10, pavlovicetal13}.

Modelling kernels based on analytical methods often require that
the underlying phenomena appear smooth and predictable. This,
however, need not be the case, especially when simplifications are
invoked to describe the complexity of the systems with a small
number of parameters. This is usually done in studies of
astrophysical objects. Evolution of such objects, despite their
complex macroscopic appearance, reflected in diversity of
intrinsic and environmental parameters, is often described with
only two parameters, as is the case with calibration relations
considered in this paper. The intrinsic complexity of the nature
itself leads to events that cannot be predicted using a simplified
analytic approach \citep{Taleb2007}. Consequently, the shape of
the data sample probability density function (PDF) might not
follow the direction of the same data sample best fit line and can
deviate from that line in Gaussian manner, even in cases of
complete and well-studied samples.

Our approach relies on numerical calculation of PDF of calibrating
data rather than on applying a fitting procedure to them. This
increases the likelihood  that the information contained in the
calibration sample is preserved and ensures  greater consistency
and more accurate calibrations.

The standard procedure to estimate PDF from data samples is to bin
the data and make histograms. The problem with this approach is
that the reconstructed PDFs can heavily depend on the bin size,
especially when incomplete calibrating samples are considered {
\citep[for more information on assessing the binning problem, we
refer the reader to][where a sophisticated approach with Bayesian
Blocks method is used]{Scargle_et_al_2013ApJ_764_167S}}. Another
possible approach is to reconstruct cumulative distribution
function (CDF) and then reconstruct the PDF \citep{BergHarris08}.
However, CDF reconstruction requires data sorting and can only be
performed on unidimensional data. Our approach requires no
histograms and no CDF reconstruction. We calculate PDF using Monte
Carlo resampling of the calibration (original) sample. Coordinates
of the data points in resampled samples are translated to account
for the difference between centroid coordinates of the original
and resampled sample. Our algorithm stems from the basic
principles of bootstrap statistics \citep[random
resampling,][]{EfronTibshirani93} and principal component analysis
\citep[standardization of data coordinates using centroids and
calculation of highest data variability direction,][]{pearson1901,
Jolliffe_PCA_2002}. Albeit simple, the algorithm is
computationally intensive. However, in present times of ubiquitous
computing resources it can be performed with sufficient accuracy
on a standard office computer and it can yield smooth PDFs that
resemble the distribution of calibration sample data points. Even
small samples (of $\sim 10 $ data points) can give smooth density
maps of high resolution (see Equation  \ref{n_resampl} and Figure
\ref{pdf_snrs_pne}). This can be very significant for calibrating
relations where scarce samples are a rule rather than exception.
We developed our algorithm for the purpose of calibrating
bidimensional data samples and our analysis and algorithm
presentation will be constrained to two dimensions.  The
simplicity of our approach makes a multidimensional data
application a quite straightforward extension (see Section 2).

We apply our analysis to the radio surface brightness to diameter
{ ($\Sigma-D$)} relation for supernova remnants (SNRs) and
planetary nebulae (PNe), and also to the period-luminosity ($PL$)
relation for classical Cepheids. The majority of the work done on
the $\Sigma-D$ relation was in order to produce reliable
calibrations that can be used to calculate distances. Samples
often suffer from significant scatter and if the information that
they contain is described with the parameters of the best fit
line, it can result in inaccurate distance estimates. On the other
hand, the $PL$ relation gives more consistent distances, with the
order of magnitude smaller average fractional error than the above
mentioned $\Sigma-D$ relations. The average fractional error for
distance is calculated as:
\begin{equation}
\bar{f}=\frac{1}{N} \sum_{i=1}^N \left| \frac{d_\mathrm{i}-d_\mathrm{i}^\mathrm{s}}{d_\mathrm{i}} \right|,
\end{equation}
where $N$ is the number of data points in the calibrating sample,
$d_\mathrm{i}$ is the measured distance to the object represented
with $\mathrm{i}^\mathrm{th}$ data point and
$d_\mathrm{i}^\mathrm{s}$ is a statistical distance to that object
determined either from the best fit line to the calibrating data
set or in some other way such as the PDF-based method presented in
this paper.

\subsection{The $\Sigma-D$ relations for supernova remnants and planetary nebulae}

The relation between the radio surface brightness and diameter is usually given as:
\begin{equation}
\Sigma = AD^{-\beta},
\label{SigD}
\end{equation}
where $A$ and $\beta$ are parameters. This is the standard form
that follows from theoretical work \citep[first derived for
supernova remnants, ][]{shklovskii60a} and is readily used for
calibration.  Calibration is performed by linearising the above
equation and applying some of the standard fitting techniques
\citep{shklovskii60b}. Also, from theoretical considerations it is
expected that $A$ and $\beta$ have different values in different
stages of SNR evolution and this can also interfere with
calibration precision when modelling SNR evolution with only one
evolutionary trajectory (the best fit line), which is usually
done. Once calibrated, the relation can be used to determine the
distance to a particular SNR by measuring its flux density $S =
\Sigma / \Omega$ and angular diameter $\theta =
\sqrt{4\Omega/\pi}$. After calculating $\Sigma$, the corresponding
value for $D$ follows from Equation \ref{SigD} and distance can be
calculated as $d=\theta D$. The $\Sigma-D$ relation for SNRs has
more than five decades' long history. In addition to further
theoretical development \citep[i.e.,][]{DuricSeaquist86},
extensive work was done on calibrating the relation for distance
determination \citep[for some calibrations
see,][]{urosevic_etal05,CaseBattacharya98,allakhverdievetal86}.
The $\Sigma-D$ relation for planetary nebulae in the form of
Equation \ref{SigD} was theoretically derived and empirically
assessed by \citet{urosevic_etal07,urosevic_etal09b}.

The above papers use standard fitting procedures based on vertical
offsets  and are mostly not concerned with further development of
fitting procedures. \citet{pavlovicetal13} argued that applying
different types of fitting offsets can result in different
parameters of the $\Sigma-D$ relation for SNRs, and that
orthogonal offsets are more reliable and stable over other types
of offsets. Similar analysis, but to a lesser extent, was
performed on a PNe sample from \citet{stanghellini_etal08} in
\citet{VukoticUrosevic12}. Although these analyses argue in favour
of orthogonal offsets calibrations, the dependence of calibration
parameters on the type of selected fitting offsets introduces
further ambiguities in the efforts towards reliable calibrations.
Also, poor quality of the calibrating samples often results in
statistically unacceptable fits.

The calibration algorithm proposed in this paper is not using
fitting procedures and there are no assumptions on the type of
functional dependence in calibrating relation. This makes the
resulting calibration more consistent, with no loss of
information, because the initial information contained in the data
points coordinates is not reduced to the parameters of the best
fit line. Here we apply our algorithm for data density
distribution calculation to the sample of 60 Galactic SNRs from
\citet{pavlovicetal13} and to the 39 Galactic PNe with reliable
distances from \citet{stanghellini_etal08}.

 \subsection{The $PL$ relation for Cepheids}
\label{intro_ceph}

In the case of Cepheids, pulsating variable stars, historically
the period-luminosity relation \citep{Leavitt_1912} was of crucial
importance for determining distances. The $PL$ relation is also a
starting point of the distance ladder \citep[e.g.][]{Rowan_1985}.
Using Cepheids to estimate distances to other galaxies is one of
the starting points in measuring the Hubble constant ($H_0$).
Small $PL$ calibration inaccuracies  can propagate to
significantly larger discrepancies in estimates of the universe
expansion rate. As the body of data increased, it became obvious
that there are some problems with $PL$ relation, some of which
remain unsolved until this day \citep{Sandage_tammann2006}. The
problems of the $PL$ relation are connected to the problem of
reddening in different direction of eyesight, investigation of the
metallicity dependence, phase dependency of the relation,
universality of the $PL$ relation, or study of the non-linearity
of the $PL$ function of various samples
\citep[e.g.][]{Garcia_Varela_2013,Kanbur_2010,Koen_2007}. The $PL$
relations in mid infra-red are somewhat less problematic than
their visual counterparts, but still, relations at $3.6~{\mu}m$
and $4.6~{\mu}m$ for samples of Large Magellanic Cloud (LMC)
Cepheids \citep[]{Freedman_2008,Ngeow_2008,scowcroft_etal_11}
leave open questions about their metallicity dependency. All of
these issues directly affect the estimation of the $H_0$.

Studies on the $PL$ relation usually present luminosity in the
form of an absolute magnitude $M$ and use $M-\log{P}$ form of the
data for plotting and fitting. Once calibrated, the $M-\log{P}$
form can be used as a distance estimation tool. Similarly to the
$\Sigma-D$ relation, if the period of a pulsating star is
measured, a corresponding $M$ value that is derived from the
parameters of the calibration (the line fitted to the calibration
sample) can be used to estimate distance $d$ to an object with a
measured value of apparent magnitude ($m_\mathrm{o}$):
\begin{equation}
d = 10^{\left( 1.0 + \frac{m_\mathrm{o}-M}{5.0}\right)}.
\label{distance_PL}
\end{equation}

We selected the LMC fundamental mode Cepheid samples in I and V
band from OGLE project \citep{Soszynski_etal2008AcA}, that was
corrected for extinction by \citet[][and references
therein]{Ngeow_etal_ApJ09}. Compared to the considered samples of
Galactic SNRs and PNe, these samples have a significantly larger
number of data points { (better plotting plane coverage)} and
should yield more accurate distance calibrations. { Also,
better accuracy is evident from the smaller scatter in the Cepheid
samples relative to the selected axis range, than in the case of
SNR and PN samples (scatter of the PDF signal from the best fit
line in Figures \ref{pdf_snrs_pne} and \ref{ceph_pdf_I_V}).
Initial conditions and host environments of SNRs and PNe are by
far more diverse than for Cepheids and consequently less accurate
when described with a single linear relation.}

This paper is organized as follows. The next section describes the
implementation of our PDF-based algorithm and its features. In
Section \ref{analysis} we present the resulting PDF for selected
samples of SNRs, PNe and Cepheids, respectively, and discuss the
results. Also, in Table 1 we give the average fractional errors of
our calibrations and compare them with previous calibrations while
our PDF-based calibrations themselves are given in Tables 2 and 3.
Summary of the results and conclusions of this paper are presented
in the last section.

\section{Method}

Rather than fitting the calibration sample we calculate the number
density distribution of data points in the 2D fitting plane --
which directly relates to probability density distribution of one
data variable for the fixed value of the remaining data variable.
Our method relies heavily on the philosophy behind bootstrap
resampling \citep{EfronTibshirani93,Pressetal07}. Instead of using
bootstrap to determine the distribution of the fit parameter
values, we used it to estimate the number density distribution of
the data points. All of the resampled data samples are plotted
against the centroid of the original sample. The coordinates of
the resampled data  $C={X,Y}$ are calculated as:
\begin{equation}
C = C_\mathrm{d} - C_\mathrm{d}^\mathrm{cnt} + C_\mathrm{r}^\mathrm{cnt},
\label{coord}
\end{equation}
where $C_\mathrm{d}$ is the corresponding coordinate of the data
point, $C_\mathrm{d}^\mathrm{cnt}$ is the corresponding coordinate
of the centroid of the original data sample, while
$C_\mathrm{r}^\mathrm{cnt}$ is the corresponding coordinate of the
centroid of the resampled data sample. Our algorithm for the
calculation of data points density distribution is as follows:
\begin{enumerate}
\item Make a rectangular grid of cells  overlayed on the plotting surface.
\item Calculate centroid coordinates\footnote{Centroid coordinates are calculated as mean values of the corresponding data sample coordinates.} for the original data sample and plot the data points.\label{2}
\item Perform a Monte Carlo resampling with repetition on the original (calibration) data sample\footnote{This is done by randomly selecting points from the calibration sample until the number of selected points equals the number of points in the sample. This being resampling with repetition, some points may be selected more than once while some might not be selected at all.} to get the resampled sample. \label{3}
\item Calculate centroid coordinates of the resampled data sample. \label{4}
\item Calculate plotting coordinates of points in the resampled data sample using Equation  \ref{coord} and plot them on the same plot as for \ref{2}. \label{5}
\item Repeat steps \ref{3}-\ref{5} as many times as necessary to get a smooth data point PDF for a given grid resolution.
\end{enumerate}

In the original bootstrap approach, random resampling is used for
pinpointing the distribution of fit parameters values. Here we
modified this approach \citep[using the SIMD-Oriented Fast
Mersenne Twister random number generator,][]{saito_matsumoto08}
for the purpose of estimating data points number density
distribution. The advantage over other mapping techniques is that
no binning and smoothing is required. Relevant parameters are
contained within the positions of the sample data points. As an
example, we present a thought experiment. If we have a sample of
several data points that are sufficiently widely spaced, after
resampling, each data point would leave a density distribution
signal in the form of a ``smudge". The shape of the ``smudge" is influenced by the shape of the whole sample, i.e. the
most elongated axis of the ``smudge" is in the direction of
the largest change in the data sample (first principal component).
Parts of the plotting plane with high data density will have an
overlap of individual ``smudges", shaping the high values
part of the data sample PDF (see Figures \ref{pdf_snrs_pne} and
\ref{ceph_pdf_I_V}). Resampling the data sample of just two points
gives only three possible resampled samples: first point sampled
twice without sampling the second point, second point sampled
twice without sampling the first point and both points sampled
once. This leaves only three possible centroids and density
distribution of poor smoothness. For a sample of $n$ data points
it is possible to have:
\begin{equation}
N^\mathrm{cnt} = \left(\frac{(2n-1)!}{n!(n-1)!}  \right)
\label{n_resampl}
\end{equation}
different centroids\footnote{This is well known in combinatorics
or multiset theory -- the number of ways that $n$ different
elements can be resampled with repetition so that each resampled
sample contains also $n$ elements.}. All of this implies that the
quality of the calculated density distribution, in terms of
smoothness and resolution, depends on the number of data points in
the given sample and their juxtaposition. In all cases we did
$10^6$ random resamplings and mapped the resulting samples on
$10^2\times10^2$ or $10^3\times10^3$ lattice spanning the
coordinates range given in plots from Figures \ref{pdf_snrs_pne}
and \ref{ceph_pdf_I_V}.

After applying the presented algorithm the resulting data sample
PDF is in the form of a 2D matrix that can be used as the
calibration for distance determination. Instead of using just the
fit line (case of fit-based calibrations), one can use the whole
PDF matrix that contains much more information about the
calibration sample than the line of the best fit to the
calibration data. If a fixed value of one variable is selected,
than one can normalize the corresponding data sample PDF matrix
row or column, in a way that it represents a PDF of the other
variable at that particular fixed value of the selected variable.
Let us denote with PDF$^{\log \Sigma=v}$, a PDF of $D$ at a fixed
value $v$ for $\log \Sigma$ and accordingly with PDF$^{\log P=v}$,
a PDF of $M$ at a fixed value $v$ for $\log P$ (in units as in
Figures \ref{pdf_snrs_pne} and \ref{ceph_pdf_I_V}).

To get a single value for the distance to a particular object,
SNR, PN, or a Cepheid, it is necessary to get a single value of
the desired property of the object from the corresponding PDF
distributions ($D$ from PDF$^{\log \Sigma=v}$ or $m$ from
PDF$^{\log P=v}$). Here, we use basic statistical properties of
these distributions for such  purpose: mode, mean and median{,
presented in Tables 2 and 3. Median is the most robust parameter
that changes very slowly with data fluctuations and represents the
middle value, with equal probability that the value of $D$, or
$m$, is situated in higher or lower values than median. This
property can be useful in assigning the error bar to the selected
diameter (statistical distance) value. Mode marks the value of the
highest probability and can be a good estimator of distance for
distributions where the corresponding mode peak dominates the
whole distribution. Although less stable than median with respect
to fluctuations in data, mean value can be useful in estimating
error and it describes the resulting PDF in more detail.}

\subsection{ Error estimates for distances to individual objects}
\label{error}

{ The estimation of error in statistical distance derived from
the PDF-based method is not trivial and straightforward, since the
method gives us a PDF distribution of the statistical distance
often not similar to any of the distributions readily used in
statistical practice. The error estimate for an individual
distance can stem from various reasons: errors in input data
values, data sample PDF estimate related errors,  errors related
to the shapes of PDF$^{\log \Sigma=v}$ or PDF$^{\log P=v}$ (i.e.
data scatter), etc. Errors related to data sample PDF estimate
should decrease as the number of resamplings increases, eventually
loosing significance compared to other sources of errors. If a
data sample PDF is shaped in a Gaussian manner (which is an
extremely rare case), the errors related to the shape of
PDF$^{\log \Sigma=v}$ (PDF$^{\log P=v}$) can be estimated as a
Gaussian standard deviation. In the case of irregular shapes, such
as the one plotted in Figure \ref{sigma_pdf}, the calculation of
error estimate may differ from case to case.

The purpose of describing the statistical distribution with an
error is to know the probability of a randomly drawn value from
that distribution to fall within the error interval. The most
frequently used error estimator is the standard deviation, which
is the expected value of the variable deviation from the mean. In
normal distribution, $68 \%$ of the values should be within one
standard deviation from the mean. For non-standard distributions
this percentage is not known in advance although it can be
constrained by means of the Bienaym\' e-Chebyshev
inequality\footnote{Probability $P$ of finding variable $x$
relative to the mean value $\mu$ at $k$ standard deviations
$\sigma$, satisfies: $P(|x-\mu| \geq k\sigma)\leq \frac{1}{k^2}$.
In example, for $k=2$, it follows that for a given PDF  one can
expect, to have at least $75\%$ probability that a data point will
fall within two standard deviations from the mean.}
\citep{Dodge08_stat_encyclop}. Instead of making constraints on
the percentage of values that are within the error interval, it is
better and more precise to calculate them numerically  from the
PDF. Here, we descriptively present two possibilities:
\begin{itemize}
\item[(a)]Let's have pointer $l$ pointing to the PDF point which is first to the left hand side of the mode point, and similarly, pointer $r$ pointing to the first next to mode point on the right hand side. Read the values of PDF at points pointed by $l$ and $r$, and denote them with $V_\mathrm{l}$ and $V_\mathrm{r}$, respectively. Increment the integral sum for the contribution of a higher value, or both values if $V_\mathrm{l} = V_\mathrm{r}$. Move each pointer that points to the value that was added to the integral sum to the next point in the same direction as before. Repeat this until the integral sum gets larger than a prescribed value, $0.7$, $0.9$, or other. Move one step back the pointer which points to a value whose contribution was not added to the integral sum. The interval between pointers $l$ and $d$ would then represent the error interval for the estimated value that encompasses the desired percentage of variable realizations. This procedure will incorporate parts of variable axis with higher PDF values in the resulting error interval and it should be well-suited for description of error intervals in PDFs with a dominant mode peak (even if wide or asymmetric). This is likely to be the case in stronger correlations.
\item[(b)]Alternatively, one can start from median value and move the pointers in each integration step to the next PDF point in their direction regardless of the $V_\mathrm{l} / V_\mathrm{r}$ ratio. This will treat in an equal manner both sides of PDF in respect to the median. Since median is a robust estimator, this algorithm might be useful (more realistic) to estimate uncertainties in cases of loose relations.
\end{itemize}
The two proposed algorithms (or their variations) for error
estimates are yet to be developed and tested. Such an extensive
work is beyond the scope of this paper and hopefully, in future
work, it will be adapted to the specific needs of particular
calibrating relations. Future studies and development of the
PDF-based calibration methods may give clues on establishing
uniform criteria for error estimates on individual distances.

Also, in a PDF-based method, the uncertainties in input values
will propagate into PDFs of different shapes. The values of all
PDFs corresponding to an input uncertainty interval should be
considered as sets of resulting means, medians and modes. The
extreme values of elements in these sets indicate the
uncertainties interval. These should be combined with the
uncertainties caused by the PDF shape, the ones that can be
obtained from algorithms (a) and (b), to calculate the final
uncertainty (error) of the distance to an individual object. }

In the next section, we demonstrate how to use Tables 2 and 3 to
get an estimate of the distance to an SNR, a PN or a Cepheid, with
a particular example of one selected Galactic SNR, {and estimate
the uncertainty of this distance by using the corresponding PDF
given in Figure \ref{sigma_pdf}.

\begin{figure}
 \includegraphics[width=0.5\textwidth]{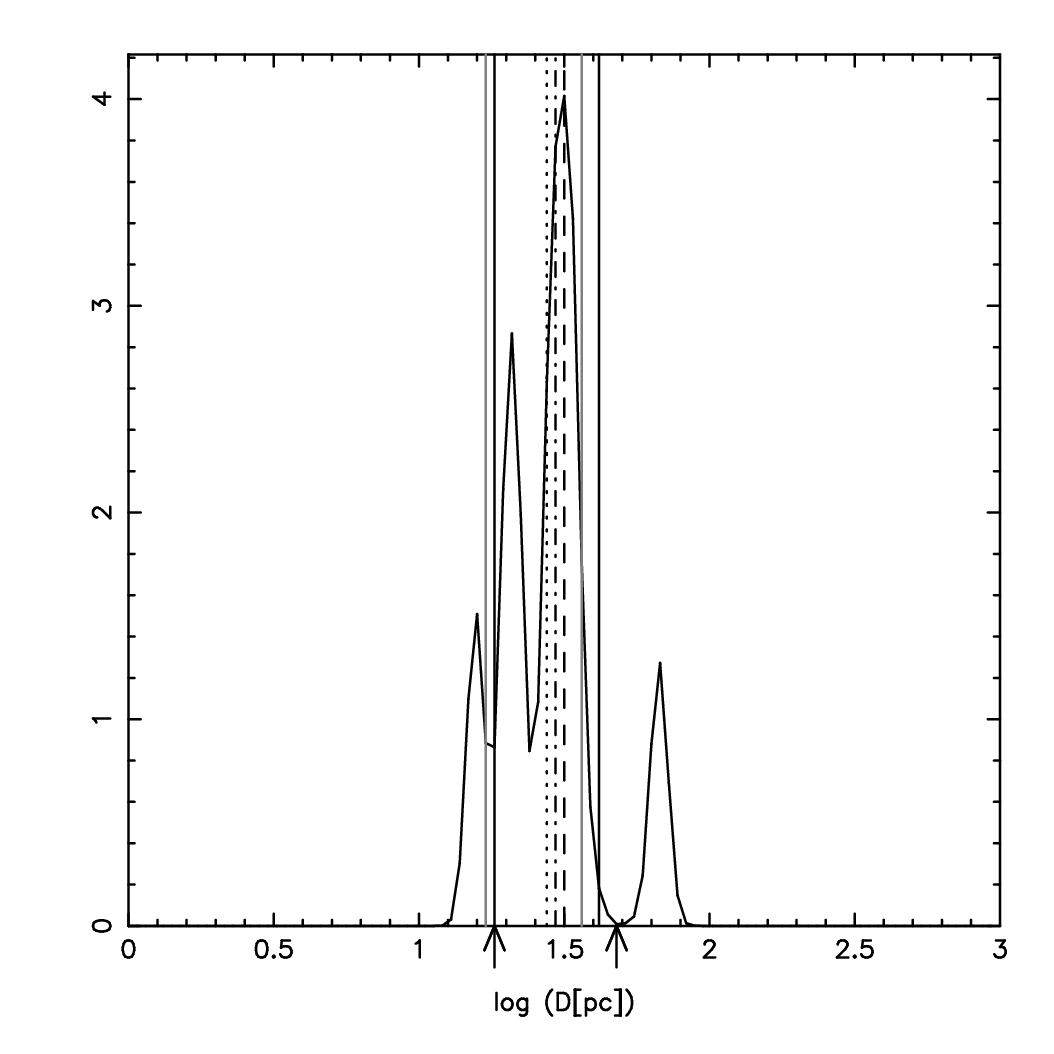}
  \caption{PDF of the diameter variable at the fixed value of $\log{\Sigma}=-19.98$ from the data sample PDF presented in the left panel of Figure \ref{pdf_snrs_pne}. Mode, median and mean are presented with dashed, dashed-dotted and dotted lines, respectively. { Vertical black solid lines mark one standard deviation confidence interval. Vertical gray lines designate $75 \%$ confidence interval around mean, biased towards higher PDF values, as explained in case (a) in Section \ref{error}. Arrows mark the $75 \%$ confidence interval symmetric around median, as stated in case (b) in Section \ref{error}.}}
 \label{sigma_pdf}
\end{figure}

\subsection{Distance estimate procedure -- the case of SNR G12.0-0.1}
\label{manual}

The calculation of distance estimate to an SNR using our
calibration tables is done in the following steps:
\begin{enumerate}
\item[1)] From $S$ and $\theta$ calculate $\log \Sigma_\mathrm{o}$ of an SNR.
\item[2)] Find the $\log \Sigma_\mathrm{t}$ value in the calibration table that is the closest match to $\log \Sigma_\mathrm{o}$.
\item[3)] Use the $D$ values of mode, mean and median (for the PDF$^{\log \Sigma_\mathrm{t}}$) from the $\log \Sigma_\mathrm{t}$ table row to calculate distance as $d=D/\theta$. If the distance estimates for mode, mean and median are close together, then mode value should be used as the most probable one. In case where these three estimates differ significantly, an inspection of PDF$^{\log \Sigma_\mathrm{t}}$ may be required, either from a data sample PDF (Figure \ref{pdf_snrs_pne}), or directly, as presented in Figure \ref{sigma_pdf}. This might be the case for a multi-modal $\log \Sigma_\mathrm{t}$ or parts of the plotting plane scarcely populated with data points.
\end{enumerate}

From values presented in Table 3 of \citet{pavlovicetal13} we
calculate for G12.0-0.1 $\log \Sigma = -19.9686$ (the
corresponding PDF$^{\log \Sigma=-19.9686}$ from data PDF in Figure
\ref{pdf_snrs_pne} is plotted in Figure \ref{sigma_pdf}). The
value of ${\log \Sigma=-19.9686}$ falls within the coordinate
range of $\log \Sigma=(-20.02, -19.94)$, centred at $\log
\Sigma=-19.98$ (corresponding row in Table 2). Similarly, at
$10^3\times10^3$ resolution (Table 3), the $\log \Sigma =
-19.9686$ corresponds to $\log \Sigma = -19.972$ row. From rows
corresponding to $\log \Sigma = -19.9686$ in Tables 2 and 3 we
read out the $D$ values for mode, mean and median parameters of
the PDF$^{\log \Sigma=-19.9686}$. Inserting $\theta=7'$, which is
the angular diameter value listed in Table 3 of
\citet{pavlovicetal13}, in $d=D\theta$, we obtain
 the distance to SNR G12.0-0.1. More precisely,
obtained values are $15.5$, $13.5$ and $14.5$ kpc for mode, mean
and median at $10^2\times10^2$ resolution and $15.2$, $13.7$ and
$14.1$ kpc at $10^3\times10^3$ resolution, respectively. In Figure
\ref{sigma_pdf}, the highest PDF peak is dominant and all three
estimators are within the range of this peak. In such a case, the
most probable value (mode) can be used with high confidence,
placing the G12.0-0.1 at the distance of $\sim15$ kpc. All three
estimates are higher than the $\sim11$ kpc estimate from
\citet{pavlovicetal13} with the most probable estimates being up
to $30\%$ higher. This value is obtained using  the same
calibrators as in \citet{pavlovicetal13} and yet giving a
significantly different result.

In the work of \citet{Yamauchi2013arXiv1309} the authors argue in
favour of association of Suzaku X-ray source J181205--1835  with a
radio-shell of G12.0-0.1 SNR. They use an X-ray spectra
absorption-column-based distance estimate and compare it with
distance estimate for G12.0-0.1 obtained from the empirical radio
$\Sigma-D$ relation from \citet{pavlovicetal13}. In this case, the
$30\%$ disagreement in $\Sigma-D$ distance estimate can mean the
difference of associating and not associating the J181205−1835
source with G12.0-0.1 SNR. When comparing Figure
\ref{pdf_snrs_pne} from this paper and Figure 1 from
\citet{pavlovicetal13}, we can see an indication of a densely
populated calibrator data point region at $\log\Sigma\sim-20.0$
which is situated to the right of the fit line. This causes the
$D$ values estimated from a PDF-based method to be higher than the
value estimated from the best line, giving larger distances for a
given angular diameter. All of this highlights the importance of
local PDF features in statistical distance calibrations that are
averaged out in the case of methods based on fitting procedures.

In the case of Cepheids, similarly, the $m$ values (mode, mean and
median) corresponding to PDF$^{\log P=v}$ in Tables 2 and 3 are used to calculate  $M = m + 5 (1 - \log {d_\mathrm{LMC}})$ which is then
substituted in Equation 3, along with the corresponding value for
$m_\mathrm{o}$ in order to calculate the distance estimate (for distance to LMC we used $d_\mathrm{LMC} = 49970.0$ pc).

{ As for the error estimates on obtained statistical distances,
at $\pm 2 \sigma$ ($\sigma$ designating the standard deviation)
interval around mean, the Bienaym\' e-Chebyshev inequality gives
that minimum $75 \%$ of the diameters should have values inside
this interval. However, this constraint is of hardly any use since
we integrated that in the case of G12.0-0.1 (Figure
\ref{sigma_pdf}) $78 \%$ of the values are already within $\pm
\sigma$ interval (vertical solid black lines in Figure
\ref{sigma_pdf}). These boundaries correspond to distances of
$8.94$ and $20.47$ kpc. Algorithm (a), in favour of higher PDF
values, gives $75 \%$ confidence interval boundaries at $8.34$ and
$17.83$ kpc, plotted with vertical solid gray lines (Figure
\ref{sigma_pdf}). The same level of confidence for a median-based
algorithm (b), plotted with arrows in Figure \ref{sigma_pdf}, is
from $8.94$ to $23.51$ kpc. As expected, interval from (a),
compared to interval from (b) and even to $\pm \sigma$ interval,
has the highest PDF value density per unit length of $\log D$ axis
and is the narrowest. Even at this, best constrained case at
$75\%$ confidence level, the distance uncertainty is $9.49$ kpc,
i.e. higher than the lower boundary of the confidence interval.
Such a low accuracy (due to a large scatter in the calibrating
sample) is of hardly any use for mapping distances, but insight in
PDF distributions (such as the one in Figure \ref{sigma_pdf}) can
be of great use for selecting a fiducial distance estimate (as
described in the previous paragraph). The future work on
developing methods for integration of confidence intervals and
analysis of their span compared to corresponding PDFs, should
hopefully result in well-developed algorithms that will be
suitable for uniform application to all fixed value PDFs in a 2D
PDF matrix and will yield confidence intervals in cases of, for
example, values given in Tables 2 and 3.}

\section{Analysis and Results}
\label{analysis}

Here we present and discuss the resulting PDFs for four selected
data samples (Figures \ref{pdf_snrs_pne} and \ref{ceph_pdf_I_V}).
The corresponding $\bar{f}$ values are given in Table 1. The
``fit" column values for $\bar{f}$ were calculated using the fit
parameters from the references in the corresponding ``Sample"
column. In the case of SNRs and PNe we selected fit parameters of
the procedure that used  orthogonal offsets from the best fit
line, while in the Cepheid samples the authors fitted the samples
with standard $Y(X)$ linear regression using the offsets along the
$Y$-axis {(this is well-suited since pulsating periods are
determined with high precision from OGLE light curves, compared to
luminosity, while in the case of $\Sigma-D$ relations for SNRs and
PNe there are significant uncertainties in both variables)}. With
the use of standard fit-based calibrations the information
contained in the data sample is condensed into the parameters of
the fit line along the underlying, fit-related statistical
assumptions. While this is helpful in terms of grasping the
general evolutionary trends of the data and shorter description of
the data sample, it can lead to substantial local inconsistencies
when used for distance determination.

Although these local deviations from the fit line are
statistically incorporated into the resulting parameters of the
calibrating fit, they are often averaged out to a significant
degree. Inspection of Figures \ref{pdf_snrs_pne} and
\ref{ceph_pdf_I_V} and Table \ref{fittable} shows that PDF-based
calibrations are better in tracing the local deviations in the
sample, resulting in smaller $\bar{f}$. It follows from Equation
\ref{n_resampl} that the number of possible different resamplings,
in the case of the sample with the smallest number of data points
(39 PNe objects), is $\sim10^{22}$. Selected number of $10^6$
resamplings gives enough lattice counts to grasp the significant
features of data sample PDF and can be computed with ease on an
average desktop computing machine of today. Changing the number of
lattice cells from $10^2\times10^2$ or $10^3\times10^3$ did not
significantly change the results (Table \ref{fittable}). The
difference between these two cases is smaller than the difference
between fit-based and PDF-based calibrations.

\begin{table*}
 \centering
 \begin{minipage}{140mm}
  \caption{Average fractional errors in distance determination for calibrating samples fitted in cited works and using the calibration method presented in this paper. Values are given in cases of $10^2\times 10^2$ and $10^3\times 10^3$ grid resolution. All values are expressed as percent.}
  \begin{tabular}{ccccccccc}
  \hline
Sample & fit & \multicolumn{3}{c}{$10^2 \times 10^2$} & &\multicolumn{3}{c}{$10^3 \times 10^3$} \\\cline{3-5}\cline{7-9}
 & & mode & mean & median & & mode & mean & median\\\cline{3-5}\cline{7-9}
\hline
SNRs, \citet{pavlovicetal13} &47.21 &33.15 &38.76 &35.70 &&30.60 &37.07 &35.49 \\
PNe, \citet{VukoticUrosevic12} &48.64 &44.13 &44.33 &43.58 && 40.42 &41.72 &43.04 \\
Ceph I, \citet{Ngeow_etal_ApJ09} &\phantom{0}4.85 &\phantom{0}4.88 &\phantom{0}4.76 &\phantom{0}4.88 &&\phantom{0}4.78 &\phantom{0}4.56 &\phantom{0}4.46 \\
Ceph V, \citet{Ngeow_etal_ApJ09} &\phantom{0}7.57 &\phantom{0}7.72 &\phantom{0}7.40 &\phantom{0}7.50 &&\phantom{0}8.00 &\phantom{0}7.17 &\phantom{0}7.01 \\
\hline
\end{tabular}
\end{minipage}
\label{fittable}
\end{table*}

\subsection{Galactic SNRs and PNe}

In Table 1 the $\bar{f}$ values in ``fit" column, taken from
\citet{pavlovicetal13}, were obtained using the  orthogonal fit
parameters. The PDF of the SNRs sample gives up to $16$ percentage points smaller
$\bar{f}$ than in the case of the best fit line. This corresponds
to the mode parameter, while median and mean give somewhat larger
$\bar{f}$. This implies that the sample is dominated by widely
scattered small subsamples of data points (left panel in Figure
\ref{pdf_snrs_pne}). {\bf ({A})} The first reason for this is that
the sample is incomplete and there are unsampled parts of the
parameter space populated with every available SNR object out
there. In this case, the scatter can be attributed to poorly
determined distances of the calibrators that scatter the data
points across $\log \Sigma-\log D$ plane. {\bf ({B})} The second
explanation is that the sample is sufficiently complete to show
out  the features of the SNRs real PDF and that diversity and
complexity of SNRs evolution cannot be faithfully described only
by using $\Sigma$ and $D$ and drawing a single fit line through
the calibrating sample. This was pointed out in
\citet{arbutina_urosevic_2005MNRAS} where authors argue that
diversity in the density of the SNR surrounding environment can
result in SNRs evolving along parallel tracks in the $\Sigma-D$
plane, according to the density of their surroundings. Also, the
total energy of the supernovae explosion can be a dominant
evolutionary agent in the early stages of SNR evolution,
eliminating dependence on the density of the surrounding matter
\citep{BerezhkoVolk2004AA}. In this scenario, the PDF-based method
for distance calibration is much more reliable than just a single
fit line-based calibration, as it gets a firmer hold on SNRs
complex evolutionary features. For example, if $\Sigma_\mathrm{o}$
of particular SNR is measured, its distance can be determined more
accurately from a PDF at fixed $\Sigma_\mathrm{o}$ along the $D$
coordinate (Figure \ref{sigma_pdf}) if the ambient density of the
measured SNR $\rho_\mathrm{o}$ can be constrained relative to the
ambient density of the calibrating SNRs. This will give clues as
to which local extreme of the example PDF from Figure
\ref{sigma_pdf} can represent the most probable track of the
evolution for SNR with $\Sigma_\mathrm{o}$ and $\rho_\mathrm{o}$
with more accurate SNR distance as a result. In order to
strengthen the reasoning from (B), it is of paramount importance
that calibrating SNR samples have accurately determined distances,
independent of the $\Sigma-D$ relation. Admittedly, this is a huge
task, but large future surveys in different parts of the
electromagnetic spectrum can give sufficient data for this goal to
become feasible.

The general cases of (A) and (B) also hold for the examined PNe
sample (Figure \ref{pdf_snrs_pne} and Table \ref{fittable}).
Accordingly, a slightly larger $\bar{f}$ than in the case of
SNR sample can be attributed either to a smaller number of data
points (A) or great diversity in PNe evolution (B). The second
case finds good grounds in the work of \citet{FrewParker2010PASA}
where authors emphasize the diversity of PN properties and
different evolutionary scenarios. As in the case of SNRs, larger
surveys and data bases will provide more reliable calibrators and
give more clues on how to calibrate the $\Sigma-D$ relation.

\begin{figure*}
 \includegraphics[width=0.45\textwidth]{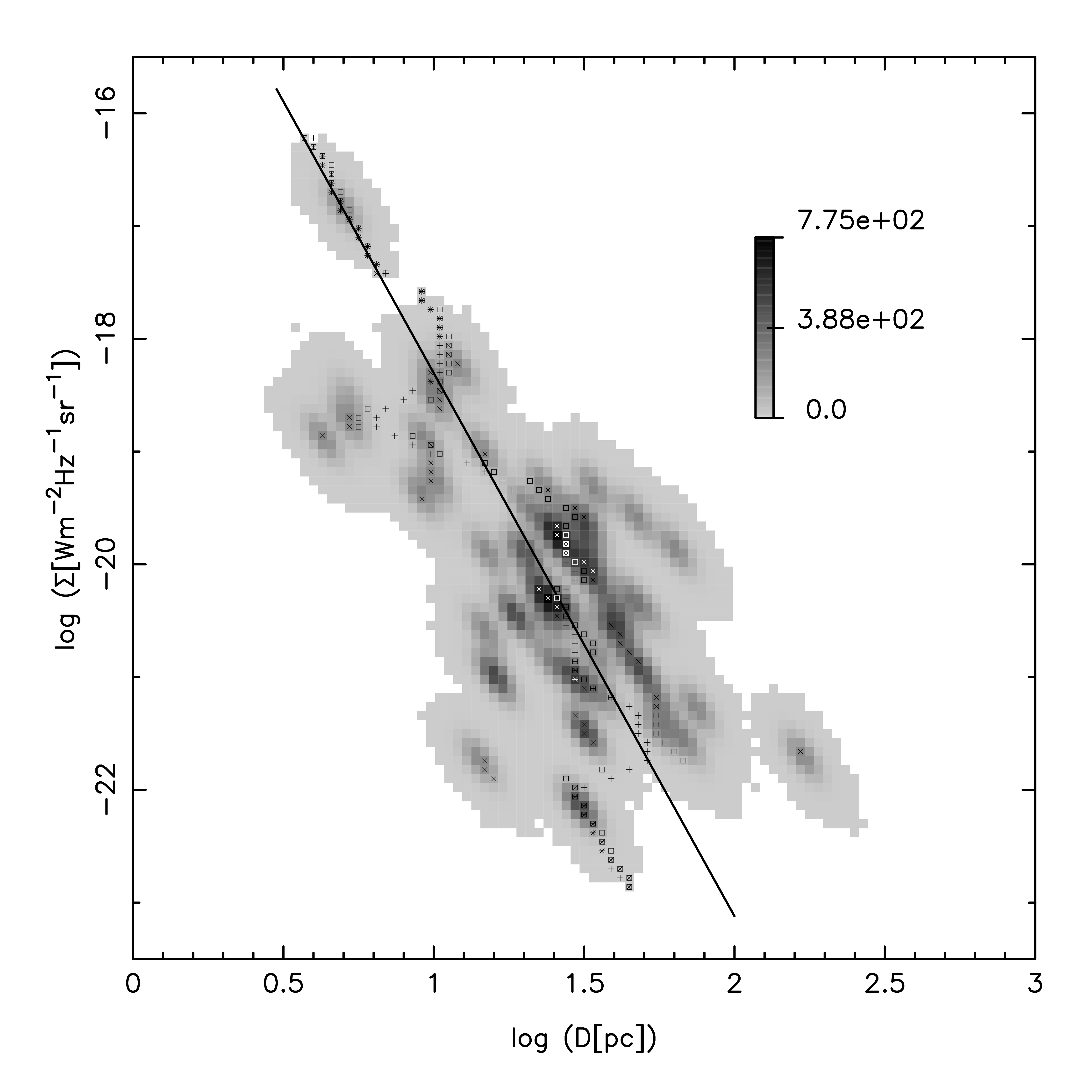}
 \includegraphics[width=0.45\textwidth]{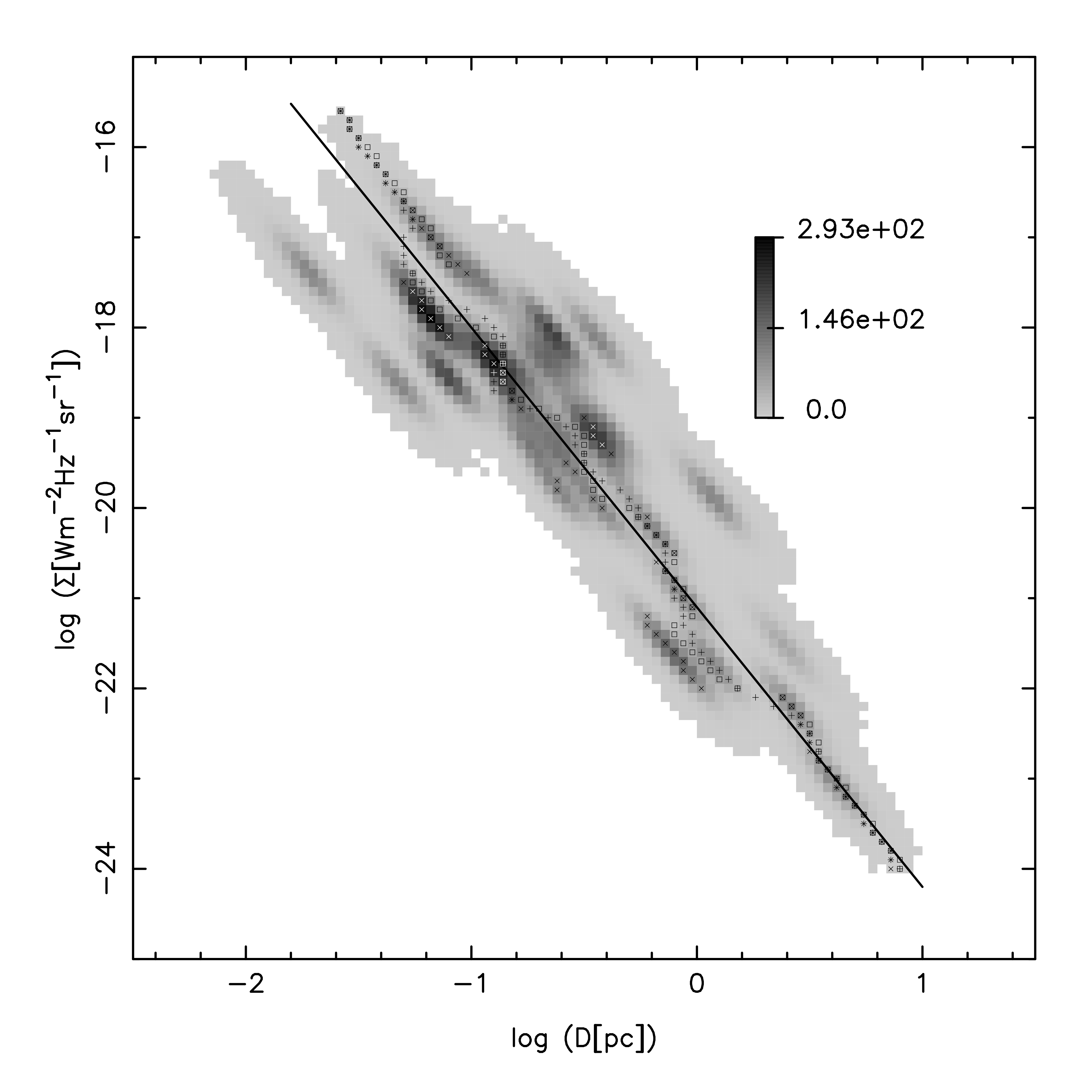}
  \caption{Grayscale reconstructed data PDF. The lattice of $100\times 100$ cells is mapped on the variables range shown on the plot.  The markers represent parameters of the distributions at fixed $\Sigma$ values, along the $D$ axis (rows of the plotted PDF matrix): mode -- diagonal cross, median -- open square and mean -- cross. Left: 60 Galactic SNRs sample from \citet{pavlovicetal13}. Right: 39 Galactic PNe sample from \citet{stanghellini_etal08}. Solid line is the orthogonal offsets best fit line with parameters from corresponding works of \citet{pavlovicetal13} and \citet{VukoticUrosevic12}.}
  \label{pdf_snrs_pne}
\end{figure*}

\subsection{LMC Cepheids}

In the case of the Cepheid samples, the PDF-based method gives
similar results as the standard fitting  method. Somewhat larger
values were obtained for estimates from mode parameter in both examined bands and median parameter in I band than for
estimates from the fitting method, but smaller $\bar{f}$ for mean parameter in both examined bands
and median parameter in V band (Table 1 and Figure
\ref{ceph_pdf_I_V}). The reduction in $\bar{f}$ for V band is
larger than in I band and is up to $\approx0.6$ percentage points (median) and
$\approx0.4$ percentage points (median). This is in agreement with arguments of
Section \ref{intro_ceph}, that the $PL$ relation in mid infra-red
has stronger correlation and is less problematic than in V band.
Compared to the samples of SNRs and PNe the Cepheid samples are
much more compact  and more accurate in terms of $\bar{f}$. This
is understandable given that all LMC Cepheids are approximately at
the same distance as the host galaxy. It follows that scatter
caused by inadequate calibrator distances is much smaller in the
case of LMC Cepheids. On the other hand, given the distance ladder
propagation of uncertainties, even small improvements in $PL$
calibration can significantly improve accuracy in determining the
Hubble parameter.

In addition to potential for providing more precise distance
estimates,  the more informative nature of the PDF method might
give some insight into additional issues of the $PL$ relation for
LMC Cepheids (the PDFs from Figure \ref{ceph_pdf_I_V}). At the
$10^2 \times 10^2$ resolution there are not as much distinctive
outlying features of the PDF as in the case of SNR and PN samples.
Other than that, smaller deviations of markers (that present
means, medians and modes at fixed $P$ values along $M$ axis) from
the fit line, and their mutual scatter, is evident. This
possibility should be more viable especially at higher grid
resolution where finer details of the PDF can emerge. Inspection
of small deviations of these details and asymmetries from the
``central ridge" of the PDF (in an ideal case scenario of the
well-fitting model with Gaussian noise, the PDF ``central ridge"
should be well-defined and symmetric), can help in resolving
standard $PL$ issues stated in Section \ref{intro_ceph} and give a
more precise insight into the theory of the pulsating nature of
these objects. However, it might be a delicate task to distinguish
the individual contribution of each of these phenomena to the
observed PDF landscape. Interpreting the fine details in  PDF
shape is beyond the scope of this paper and is left for a future
study.

\begin{figure*}
  \includegraphics[width=0.45\textwidth]{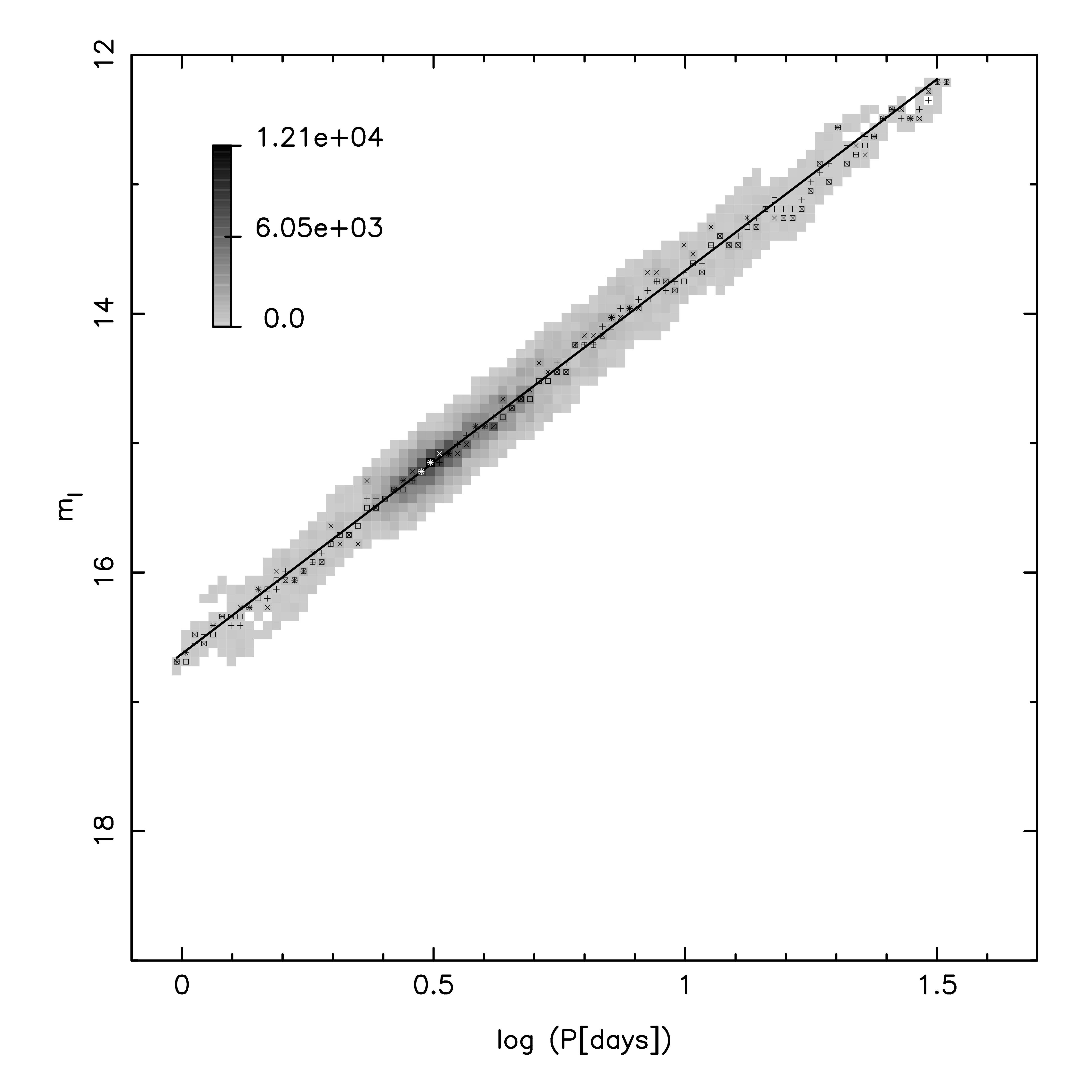}
  \includegraphics[width=0.45\textwidth]{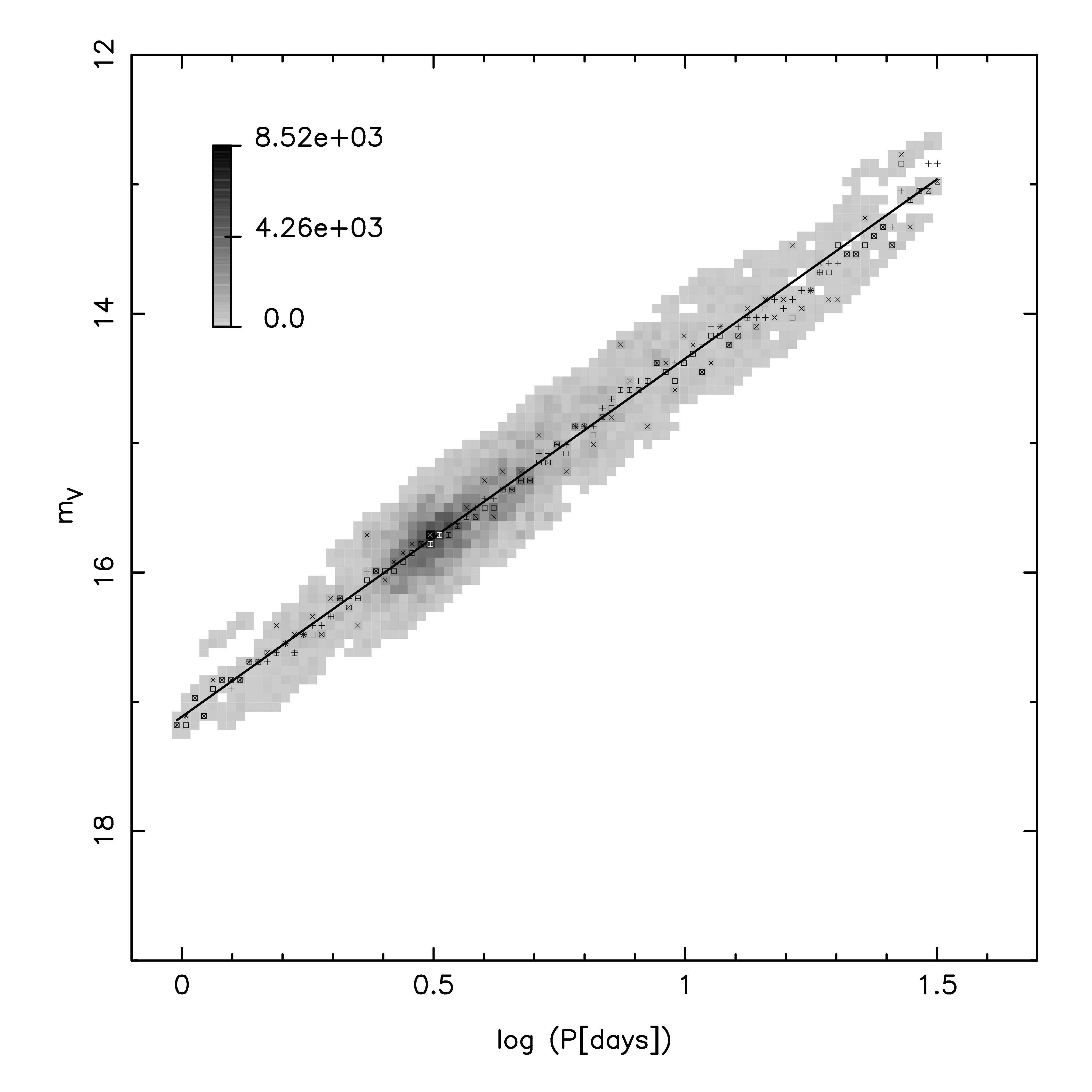}
  \caption{Grayscale reconstructed data PDF for an extinction corrected sample of fundamental mode LMC Cepheids from  \citet{Ngeow_etal_ApJ09}. The lattice of $100\times 100$ cells is mapped on the variables range shown on the plot. The markers represent parameters of the distributions at fixed $P$ values, along the $m$ axis (columns of the plotted PDF matrix): mode -- diagonal cross, median -- open square and mean -- cross. Left: 1649 objects in I band. Right: 1675 objects in V band. Solid line is the vertical offsets best fit line with parameters from corresponding work of \citet{Ngeow_etal_ApJ09}.}
   \label{ceph_pdf_I_V}
\end{figure*}

Finally, in Tables 2 and 3 we give distance calibrations from the
presented PDF-based method for all three estimators (mode, mean
and median) in selected samples of SNRs, PNe and Cepheids. As
described in Section \ref{manual}, these calibration tables can be
used to estimate distances.

\section{Summary and Conclusions}
We presented a calibration method which relies on density
distribution of data points rather than fitting procedures. The
resulting calibrations are more robust and accurate and require no
assumptions on functional dependence such as fitting-based
calibrations. Our algorithm for generating data sample PDF is
based on calculating centroid offsets between Monte Carlo
resampled and the actual calibrating data sample. As such, it
requires no binning such as in histogram-based approaches. The
method is applied to distance related scaling relations, the
$\Sigma-D$ relation for SNRs and PNe and the $PL$ relation for
Cepheids. The selected samples of Galactic SNRs and PNe have a
much larger scatter than selected samples of fundamental mode LMC
Cepheids in I and V band. This is due to the fact that Galactic
distance mapping is tenuous compared to the case where all objects
in the sample reside in external galaxy and are approximately at
the same distance.  Compared to the best calibrating fit lines for
the selected SNRs and PNe, our method gives up to $\approx 16$
and $\approx 8$ percentage points, respectively, smaller average fractional error
for distance estimates. Even in the case of LMC Cepheids, a mere
fractional error reduction of up to $\approx 0.5$ percentage points can be a
significant improvement in building up a Cepheid-based distance
ladder.

Apart from improvement in accuracy, the PDF-based calibrating
method presented in this paper gives much more information about
the data sample than standard fitting techniques. In the proposed
method, the information contained in the calibrating samples is
preserved rather than averaged out and condensed into the
parameters of the best fit line. In the case where data samples
are reliable and complete, this preserved information can be used
to give more insight into the evolution of the examined objects.
This could be a viable tool for quantifying the dependence of the
Galactic SNRs evolution on the ambient medium density or some
other relevant feature and the same holds for Galactic PNe. In the
case of LMC Cepheids, small deviations in a sample PDF can be used
to trace the fine details of the nature of pulsations and
improvement of the distance ladder.

The main purpose of this paper was to present a more accurate
method for statistical distance determination. The more
informative nature of the method opens new vistas in giving clues
on the evolution of different objects through quantification  of
sample PDF features. We leave further development of the method in
this direction for future work.

\begin{table*}
\label{calibtable}
 \centering
 \begin{minipage}{170mm}
  \caption{Calibrating values for distance determination for the Galactic $\Sigma-D$ relations (SNRs and PNe) and the $PL$ relation (in the form of apparent magnitude $m$ vs. period $P$) for fundamental mode Cepheids in I and V band. The units of the $\log$ values and coordinate range are the same as in Figures \ref{pdf_snrs_pne} and \ref{ceph_pdf_I_V}. The grid size is $10^2 \times 10^2$.}
  \begin{tabular}{ccccccccccccc@{\hspace*{-0.1mm}}ccc}
  \hline
\multicolumn{4}{c}{--------------- SNRs ---------------}&\multicolumn{4}{c}{--------------- PNe ---------------}&\multicolumn{8}{c}{----------------------------- Cepheids -----------------------------}\\
$\log{\Sigma}$ &\multicolumn{3}{c}{$D[pc]$}&$\log{\Sigma}$&\multicolumn{3}{c}{$D[pc]$}&$\log{P}$&\multicolumn{3}{c}{$m_{\mathrm{I}}$}&&\multicolumn{3}{c}{$m_{\mathrm{V}}$}\\\cline{2-4}\cline{6-8}\cline{10-12}\cline{14-16}
          &mode&mean&med.&&mode&mean&med.&&mode&mean&med.&&mode&mean&med.\\\cline{2-4}\cline{6-8}\cline{10-12}\cline{14-16}
   \hline
-23.42&  --  &  --  &  --  &    -24.90&  --  &  --  &  --  &     -0.08&  --  &  --  &  --  &    &  --  &  --  &  --  \\
-23.34&  --  &  --  &  --  &    -24.80&  --  &  --  &  --  &     -0.06&  --  &  --  &  --  &    &  --  &  --  &  --  \\
-23.26&  --  &  --  &  --  &    -24.70&  --  &  --  &  --  &     -0.05&  --  &  --  &  --  &    &  --  &  --  &  --  \\
-23.18&  --  &  --  &  --  &    -24.60&  --  &  --  &  --  &     -0.03&  --  &  --  &  --  &    &  --  &  --  &  --  \\
-23.10&  --  &  --  &  --  &    -24.50&  --  &  --  &  --  &     -0.01& 16.69& 16.69& 16.69&    & 17.18& 17.18& 17.18\\
-23.02&  --  &  --  &  --  &    -24.40&  --  &  --  &  --  &      0.01& 16.62& 16.62& 16.69&    & 17.11& 17.11& 17.18\\
-22.94&  --  &  --  &  --  &    -24.30&  --  &  --  &  --  &      0.03& 16.48& 16.55& 16.48&    & 16.97& 17.04& 16.97\\
-22.86& 44.67& 44.67& 44.67&    -24.20&  --  &  --  &  --  &      0.04& 16.55& 16.48& 16.55&    & 17.11& 17.04& 17.11\\
-22.78& 44.67& 41.69& 44.67&    -24.10&  --  &  --  &  --  &      0.06& 16.41& 16.41& 16.48&    & 16.83& 16.83& 16.90\\
-22.70& 41.69& 38.90& 41.69&    -24.00&  7.24&  7.94&  7.94&      0.08& 16.34& 16.34& 16.34&    & 16.83& 16.83& 16.83\\
-22.62& 38.90& 38.90& 38.90&    -23.90&  7.24&  7.24&  7.94&      0.10& 16.34& 16.41& 16.34&    & 16.83& 16.90& 16.83\\
-22.54& 36.31& 36.31& 38.90&    -23.80&  7.24&  7.24&  7.24&      0.12& 16.27& 16.41& 16.34&    & 16.83& 16.83& 16.83\\
-22.46& 36.31& 36.31& 36.31&    -23.70&  6.61&  6.61&  6.61&      0.13& 16.27& 16.27& 16.27&    & 16.69& 16.69& 16.69\\
-22.38& 33.88& 33.88& 36.31&    -23.60&  6.03&  6.03&  6.03&      0.15& 16.13& 16.13& 16.20&    & 16.69& 16.69& 16.69\\
-22.30& 33.88& 33.88& 33.88&    -23.50&  5.50&  5.50&  6.03&      0.17& 16.27& 16.20& 16.13&    & 16.62& 16.69& 16.62\\
-22.22& 31.62& 31.62& 31.62&    -23.40&  5.50&  5.50&  5.50&      0.19& 15.99& 16.13& 16.06&    & 16.41& 16.62& 16.62\\
-22.14& 31.62& 31.62& 31.62&    -23.30&  5.01&  5.01&  5.01&      0.21& 16.06& 15.99& 16.06&    & 16.55& 16.55& 16.55\\
-22.06& 29.51& 29.51& 29.51&    -23.20&  4.57&  4.57&  4.57&      0.22& 16.06& 16.06& 16.06&    & 16.48& 16.62& 16.62\\
-21.98& 29.51& 31.62& 29.51&    -23.10&  4.17&  4.17&  4.57&      0.24& 15.99& 15.99& 15.99&    & 16.48& 16.48& 16.48\\
-21.90& 15.85& 38.90& 27.54&    -23.00&  4.17&  4.17&  4.17&      0.26& 15.85& 15.92& 15.92&    & 16.34& 16.41& 16.48\\
-21.82& 14.79& 44.67& 36.31&    -22.90&  3.80&  3.80&  3.80&      0.28& 15.92& 15.85& 15.92&    & 16.48& 16.41& 16.48\\
-21.74& 14.79& 51.29& 67.61&    -22.80&  3.47&  3.47&  3.47&      0.30& 15.64& 15.78& 15.78&    & 16.20& 16.34& 16.34\\
-21.66&165.96& 51.29& 63.10&    -22.70&  3.16&  3.47&  3.47&      0.31& 15.78& 15.71& 15.71&    & 16.20& 16.20& 16.20\\
-21.58& 33.88& 51.29& 58.88&    -22.60&  3.16&  3.16&  3.47&      0.33& 15.71& 15.64& 15.71&    & 16.27& 16.20& 16.27\\
-21.50& 31.62& 47.86& 54.95&    -22.50&  3.16&  3.16&  3.16&      0.35& 15.78& 15.64& 15.64&    & 16.41& 16.20& 16.20\\
-21.42& 31.62& 47.86& 54.95&    -22.40&  2.88&  2.88&  3.16&      0.37& 15.29& 15.43& 15.50&    & 15.71& 15.99& 16.06\\
-21.34& 29.51& 47.86& 54.95&    -22.30&  2.88&  2.63&  2.88&      0.39& 15.50& 15.43& 15.50&    & 15.99& 15.99& 15.99\\
-21.26& 54.95& 44.67& 54.95&    -22.20&  2.63&  2.19&  2.63&      0.40& 15.43& 15.43& 15.43&    & 16.06& 15.99& 15.99\\
-21.18& 54.95& 38.90& 38.90&    -22.10&  2.40&  1.82&  2.40&      0.42& 15.36& 15.36& 15.36&    & 15.92& 15.92& 15.99\\
-21.10& 31.62& 33.88& 33.88&    -22.00&  1.05&  1.51&  1.51&      0.44& 15.29& 15.29& 15.36&    & 15.85& 15.85& 15.92\\
-21.02& 29.51& 29.51& 31.62&    -21.90&  0.95&  1.38&  1.26&      0.46& 15.22& 15.29& 15.29&    & 15.78& 15.85& 15.85\\
-20.94& 29.51& 29.51& 29.51&    -21.80&  0.87&  1.26&  1.15&      0.48& 15.22& 15.22& 15.22&    & 15.78& 15.78& 15.78\\
-20.86& 47.86& 29.51& 29.51&    -21.70&  0.87&  1.15&  1.05&      0.49& 15.15& 15.15& 15.15&    & 15.71& 15.78& 15.78\\
-20.78& 44.67& 29.51& 33.88&    -21.60&  0.79&  1.05&  0.95&      0.51& 15.08& 15.15& 15.15&    & 15.71& 15.71& 15.71\\
-20.70& 41.69& 29.51& 33.88&    -21.50&  0.72&  0.95&  0.87&      0.53& 15.08& 15.08& 15.08&    & 15.64& 15.71& 15.71\\
-20.62& 41.69& 29.51& 31.62&    -21.40&  0.66&  0.95&  0.79&      0.55& 15.08& 15.01& 15.08&    & 15.64& 15.64& 15.64\\
-20.54& 38.90& 27.54& 29.51&    -21.30&  0.60&  0.87&  0.79&      0.57& 15.01& 14.94& 15.01&    & 15.50& 15.57& 15.57\\
-20.46& 25.70& 27.54& 27.54&    -21.20&  0.60&  0.87&  0.95&      0.58& 14.87& 14.87& 14.94&    & 15.57& 15.50& 15.57\\
-20.38& 25.70& 27.54& 27.54&    -21.10&  0.95&  0.87&  0.95&      0.60& 14.87& 14.87& 14.87&    & 15.29& 15.43& 15.50\\
-20.30& 23.99& 27.54& 25.70&    -21.00&  0.87&  0.79&  0.87&      0.62& 14.87& 14.80& 14.87&    & 15.57& 15.43& 15.50\\
-20.22& 22.39& 27.54& 25.70&    -20.90&  0.79&  0.79&  0.87&      0.64& 14.66& 14.73& 14.80&    & 15.22& 15.36& 15.36\\
-20.14& 33.88& 29.51& 31.62&    -20.80&  0.79&  0.79&  0.79&      0.66& 14.73& 14.73& 14.73&    & 15.36& 15.36& 15.36\\
-20.06& 33.88& 29.51& 31.62&    -20.70&  0.72&  0.72&  0.72&      0.67& 14.66& 14.66& 14.66&    & 15.22& 15.29& 15.29\\
-19.98& 31.62& 27.54& 29.51&    -20.60&  0.66&  0.72&  0.79&      0.69& 14.59& 14.59& 14.66&    & 15.29& 15.29& 15.29\\
-19.90& 27.54& 27.54& 27.54&    -20.50&  0.79&  0.72&  0.79&      0.71& 14.38& 14.52& 14.52&    & 14.94& 15.08& 15.15\\
-19.82& 27.54& 27.54& 27.54&    -20.40&  0.72&  0.72&  0.72&      0.73& 14.45& 14.45& 14.52&    & 15.15& 15.08& 15.15\\
-19.74& 25.70& 27.54& 27.54&    -20.30&  0.66&  0.66&  0.66&      0.75& 14.45& 14.38& 14.45&    & 15.01& 15.01& 15.01\\
-19.66& 25.70& 27.54& 27.54&    -20.20&  0.60&  0.60&  0.60&      0.76& 14.45& 14.38& 14.45&    & 15.22& 15.01& 15.08\\
-19.58& 31.62& 27.54& 29.51&    -20.10&  0.60&  0.55&  0.55&      0.78& 14.24& 14.24& 14.24&    & 14.87& 14.87& 14.87\\
-19.50& 29.51& 23.99& 27.54&    -20.00&  0.38&  0.55&  0.50&      0.80& 14.17& 14.24& 14.24&    & 14.87& 14.87& 14.87\\
-19.42&  9.12& 20.89& 23.99&    -19.90&  0.35&  0.50&  0.38&      0.82& 14.17& 14.24& 14.24&    & 15.01& 14.87& 14.94\\
-19.34& 23.99& 18.20& 22.39&    -19.80&  0.24&  0.46&  0.35&      0.84& 14.17& 14.10& 14.17&    & 14.80& 14.73& 14.80\\
-19.26&  9.77& 16.98& 20.89&    -19.70&  0.24&  0.38&  0.35&      0.85& 14.03& 14.03& 14.10&    & 14.80& 14.66& 14.73\\
-19.18&  9.77& 14.79& 15.85&    -19.60&  0.29&  0.35&  0.32&      0.87& 14.03& 13.96& 14.03&    & 14.24& 14.59& 14.59\\
-19.10&  9.77& 12.88& 14.79&    -19.50&  0.26&  0.32&  0.32&      0.89& 13.96& 13.96& 13.96&    & 14.52& 14.59& 14.59\\

\hline
\end{tabular}
\end{minipage}
\end{table*}

\addtocounter{table}{-1}
\begin{table*}
 \centering
 \begin{minipage}{170mm}
  \caption{-- {\it continued}}
  \begin{tabular}{ccccccccccccc@{\hspace*{-0.1mm}}ccc}
  \hline
\multicolumn{4}{c}{--------------- SNRs ---------------}&\multicolumn{4}{c}{--------------- PNe ---------------}&\multicolumn{8}{c}{----------------------------- Cepheids -----------------------------}\\
  $\log{\Sigma}$ &\multicolumn{3}{c}{$D[pc]$}&$\log{\Sigma}$&\multicolumn{3}{c}{$D[pc]$}&$\log{P}$&\multicolumn{3}{c}{$m_{\mathrm{I}}$}&&\multicolumn{3}{c}{$m_{\mathrm{V}}$}\\\cline{2-4}\cline{6-8}\cline{10-12}\cline{14-16}
          &mode&mean&med.&&mode&mean&med.&&mode&mean&med.&&mode&mean&med.\\\cline{2-4}\cline{6-8}\cline{10-12}\cline{14-16}
   \hline
-19.02& 14.79&  9.77& 10.47&    -19.40&  0.42&  0.32&  0.32&      0.91& 13.96& 13.89& 13.96&    & 14.59& 14.52& 14.59\\
-18.94&  9.77&  8.51&  9.77&    -19.30&  0.38&  0.29&  0.32&      0.93& 13.68& 13.82& 13.89&    & 14.87& 14.52& 14.52\\
-18.86&  4.27&  7.41&  8.51&    -19.20&  0.35&  0.29&  0.32&      0.94& 13.68& 13.75& 13.75&    & 14.38& 14.38& 14.38\\
-18.78&  5.25&  6.46&  5.62&    -19.10&  0.35&  0.26&  0.29&      0.96& 13.75& 13.82& 13.75&    & 14.38& 14.45& 14.45\\
-18.70&  5.25&  6.46&  5.62&    -19.00&  0.32&  0.22&  0.24&      0.98& 13.82& 13.75& 13.82&    & 14.59& 14.38& 14.52\\
-18.62& 10.47&  6.92&  6.03&    -18.90&  0.17&  0.18&  0.20&      1.00& 13.47& 13.68& 13.75&    & 14.17& 14.38& 14.38\\
-18.54& 10.47&  7.94&  9.77&    -18.80&  0.15&  0.15&  0.17&      1.02& 13.54& 13.61& 13.61&    & 14.24& 14.31& 14.31\\
-18.46& 10.47&  8.51& 10.47&    -18.70&  0.15&  0.13&  0.15&      1.03& 13.68& 13.61& 13.68&    & 14.45& 14.24& 14.45\\
-18.38&  9.77&  9.77& 10.47&    -18.60&  0.14&  0.13&  0.14&      1.05& 13.33& 13.47& 13.47&    & 14.38& 14.10& 14.17\\
-18.30&  9.77& 10.47& 11.22&    -18.50&  0.14&  0.13&  0.14&      1.07& 13.40& 13.40& 13.40&    & 14.10& 14.10& 14.17\\
-18.22& 12.02& 10.47& 11.22&    -18.40&  0.13&  0.14&  0.14&      1.09& 13.47& 13.47& 13.47&    & 14.24& 14.24& 14.24\\
-18.14& 11.22& 10.47& 11.22&    -18.30&  0.11&  0.14&  0.14&      1.11& 13.47& 13.40& 13.47&    & 14.17& 14.10& 14.17\\
-18.06& 11.22& 10.47& 11.22&    -18.20&  0.11&  0.14&  0.14&      1.12& 13.26& 13.26& 13.33&    & 13.96& 14.03& 14.03\\
-17.98& 10.47& 10.47& 11.22&    -18.10&  0.08&  0.14&  0.13&      1.14& 13.33& 13.26& 13.33&    & 14.10& 14.03& 14.10\\
-17.90& 10.47& 10.47& 10.47&    -18.00&  0.07&  0.13&  0.10&      1.16& 13.19& 13.19& 13.19&    & 13.89& 14.03& 13.96\\
-17.82& 10.47& 10.47& 10.47&    -17.90&  0.07&  0.11&  0.09&      1.18& 13.26& 13.19& 13.12&    & 14.03& 13.89& 13.89\\
-17.74&  9.77&  9.77& 10.47&    -17.80&  0.06&  0.10&  0.07&      1.20& 13.26& 13.19& 13.26&    & 13.89& 13.96& 13.89\\
-17.66&  9.12&  9.12&  9.12&    -17.70&  0.06&  0.08&  0.07&      1.21& 13.26& 13.19& 13.26&    & 13.47& 13.89& 14.03\\
-17.58&  9.12&  9.12&  9.12&    -17.60&  0.05&  0.07&  0.06&      1.23& 13.19& 13.12& 13.19&    & 13.96& 13.82& 13.96\\
-17.50&  --  &  --  &  --  &    -17.50&  0.05&  0.06&  0.05&      1.25& 13.05& 12.98& 13.05&    & 13.82& 13.82& 13.82\\
-17.42&  6.46&  6.92&  6.92&    -17.40&  0.10&  0.05&  0.05&      1.27& 12.84& 12.91& 12.84&    & 13.61& 13.68& 13.68\\
-17.34&  6.46&  6.46&  6.46&    -17.30&  0.09&  0.05&  0.08&      1.29& 12.98& 12.84& 12.98&    & 13.89& 13.61& 13.68\\
-17.26&  6.03&  6.03&  6.03&    -17.20&  0.08&  0.05&  0.07&      1.30& 12.56& 12.56& 12.56&    & 13.89& 13.61& 13.47\\
-17.18&  6.03&  6.03&  6.03&    -17.10&  0.07&  0.05&  0.07&      1.32& 12.84& 12.70& 12.84&    & 13.54& 13.47& 13.54\\
-17.10&  5.62&  5.62&  5.62&    -17.00&  0.07&  0.05&  0.07&      1.34& 12.70& 12.77& 12.77&    & 13.54& 13.40& 13.54\\
-17.02&  5.62&  5.62&  5.62&    -16.90&  0.06&  0.05&  0.07&      1.36& 12.77& 12.63& 12.70&    & 13.26& 13.40& 13.47\\
-16.94&  5.25&  5.25&  5.25&    -16.80&  0.05&  0.05&  0.06&      1.38& 12.63& 12.63& 12.63&    & 13.40& 13.33& 13.40\\
-16.86&  4.90&  4.90&  5.25&    -16.70&  0.05&  0.05&  0.05&      1.39& 12.49& 12.49& 12.49&    & 13.33& 13.33& 13.33\\
-16.78&  4.90&  4.90&  4.90&    -16.60&  0.05&  0.05&  0.05&      1.41& 12.42& 12.42& 12.42&    & 13.47& 13.33& 13.47\\
-16.70&  4.57&  4.57&  4.90&    -16.50&  0.05&  0.05&  0.05&      1.43& 12.42& 12.49& 12.42&    & 12.77& 13.05& 12.84\\
-16.62&  4.57&  4.57&  4.57&    -16.40&  0.04&  0.04&  0.05&      1.45& 12.49& 12.49& 12.49&    & 13.33& 13.12& 13.12\\
-16.54&  4.57&  4.57&  4.57&    -16.30&  0.04&  0.04&  0.04&      1.47& 12.49& 12.42& 12.49&    & 13.05& 13.05& 13.05\\
-16.46&  4.27&  4.27&  4.57&    -16.20&  0.04&  0.04&  0.04&      1.48& 12.28& 12.35& 12.28&    & 13.05& 12.84& 13.05\\
-16.38&  4.27&  4.27&  4.27&    -16.10&  0.03&  0.03&  0.04&      1.50& 12.21& 12.21& 12.21&    & 12.98& 12.84& 12.98\\
-16.30&  3.98&  3.98&  3.98&    -16.00&  0.03&  0.03&  0.03&      1.52& 12.21& 12.21& 12.21&    &  --  &  --  &  --  \\
-16.22&  3.72&  3.98&  3.72&    -15.90&  0.03&  0.03&  0.03&      1.54&  --  &  --  &  --  &    &  --  &  --  &  --  \\
-16.14&  --  &  --  &  --  &    -15.80&  0.03&  0.03&  0.03&      1.56&  --  &  --  &  --  &    &  --  &  --  &  --  \\
-16.06&  --  &  --  &  --  &    -15.70&  0.03&  0.03&  0.03&      1.57&  --  &  --  &  --  &    &  --  &  --  &  --  \\
-15.98&  --  &  --  &  --  &    -15.60&  0.03&  0.03&  0.03&      1.59&  --  &  --  &  --  &    &  --  &  --  &  --  \\
-15.90&  --  &  --  &  --  &    -15.50&  --  &  --  &  --  &      1.61&  --  &  --  &  --  &    &  --  &  --  &  --  \\
-15.82&  --  &  --  &  --  &    -15.40&  --  &  --  &  --  &      1.63&  --  &  --  &  --  &    &  --  &  --  &  --  \\
-15.74&  --  &  --  &  --  &    -15.30&  --  &  --  &  --  &      1.65&  --  &  --  &  --  &    &  --  &  --  &  --  \\
-15.66&  --  &  --  &  --  &    -15.20&  --  &  --  &  --  &      1.66&  --  &  --  &  --  &    &  --  &  --  &  --  \\
-15.58&  --  &  --  &  --  &    -15.10&  --  &  --  &  --  &      1.68&  --  &  --  &  --  &    &  --  &  --  &  --  \\
\hline
\end{tabular}
\end{minipage}
\end{table*}

\begin{table*}
\label{calibtable2}
 \centering
 \begin{minipage}{170mm}
  \caption{Calibrating values for distance determination for the Galactic $\Sigma-D$ relations (SNRs and PNe) and the $PL$ relation (in the form of apparent magnitude $m$ vs. period $P$) for fundamental mode Cepheids in I and V band. The units of the $\log$ values and coordinate range are the same as in Figures \ref{pdf_snrs_pne} and \ref{ceph_pdf_I_V}. The grid size is $10^3 \times 10^3$. The complete table is available as an on-line material.}
  \begin{tabular}{ccccccccccccc@{\hspace*{-0.1mm}}ccc}
  \hline
\multicolumn{4}{c}{--------------- SNRs ---------------}&\multicolumn{4}{c}{--------------- PNe ---------------}&\multicolumn{8}{c}{----------------------------- Cepheids -----------------------------}\\
$\log{\Sigma}$ &\multicolumn{3}{c}{$D[pc]$}&$\log{\Sigma}$&\multicolumn{3}{c}{$D[pc]$}&$\log{P}$&\multicolumn{3}{c}{$m_{\mathrm{I}}$}&&\multicolumn{3}{c}{$m_{\mathrm{V}}$}\\\cline{2-4}\cline{6-8}\cline{10-12}\cline{14-16}
          &mode&mean&med.&&mode&mean&med.&&mode&mean&med.&&mode&mean&med.\\\cline{2-4}\cline{6-8}\cline{10-12}\cline{14-16}
   \hline
-23.492&  --  &  --  &  --  &   -24.990&  --  &  --  &  --  &    -0.0982&  --  &  --  &  --  &  &  --  &  --  &  --  \\
-23.484&  --  &  --  &  --  &   -24.980&  --  &  --  &  --  &    -0.0964&  --  &  --  &  --  &  &  --  &  --  &  --  \\
-23.476&  --  &  --  &  --  &   -24.970&  --  &  --  &  --  &    -0.0946&  --  &  --  &  --  &  &  --  &  --  &  --  \\
-23.468&  --  &  --  &  --  &   -24.960&  --  &  --  &  --  &    -0.0928&  --  &  --  &  --  &  &  --  &  --  &  --  \\
-23.460&  --  &  --  &  --  &   -24.950&  --  &  --  &  --  &    -0.0910&  --  &  --  &  --  &  &  --  &  --  &  --  \\
-23.452&  --  &  --  &  --  &   -24.940&  --  &  --  &  --  &    -0.0892&  --  &  --  &  --  &  &  --  &  --  &  --  \\
-23.444&  --  &  --  &  --  &   -24.930&  --  &  --  &  --  &    -0.0874&  --  &  --  &  --  &  &  --  &  --  &  --  \\
-23.436&  --  &  --  &  --  &   -24.920&  --  &  --  &  --  &    -0.0856&  --  &  --  &  --  &  &  --  &  --  &  --  \\
-23.428&  --  &  --  &  --  &   -24.910&  --  &  --  &  --  &    -0.0838&  --  &  --  &  --  &  &  --  &  --  &  --  \\
\hline
\end{tabular}
\end{minipage}
\end{table*}

\section{Acknowledgments}

The authors would like to thank Chow-Choong Ngeow for providing
the Cepheid data, the anonymous referee for insightful comments
that have greatly improved the quality of the manuscript and
Dragana Momi\'c for language assistance. B.V. acknowledges
inspiring discussions with Milan M. \'Cirkovi\'c, Srdjan
Samurovi\'c, Zoran Kne\v zevi\'c and Ana Vudragovi\'c. The authors
also acknowledge financial support from the Ministry of Education,
Science and Technological Development of the Republic of Serbia
through the projects 176004, 176005 and 176021. This research has
made use of NASA's Astrophysics Data System.

\bibliographystyle{mn2e}
        \bibliography{mybib}

\end{document}